\documentclass[aps,prb,twocolumn,showpacs]{revtex4}
\usepackage{graphicx}
\usepackage{natbib}

\begin{document}

\title{Particle rearrangements during transitions between local
minima of the potential energy landscape of a supercooled
Lennard-Jones liquid}

\author{Michael Vogel}
\affiliation{Department of Chemical Engineering, University of
Michigan, \\ 2300 Hayward, Ann Arbor, MI, 48109, USA}
\author{Burkhard Doliwa}
\affiliation{Max Planck Institute for Polymer Research, \\Postfach
3148, 55021 Mainz, Germany}
\author{Andreas Heuer}
\affiliation{Institute of Physical Chemistry, University of
M\"unster, \\ Schlossplatz 4/7, 48149 M\"unster, Germany}
\author{Sharon C.\ Glotzer}
\affiliation{Departments of Chemical Engineering and
\\Materials Science and Engineering, University of Michigan,
\\ 2300 Hayward, Ann Arbor, MI, 48109, USA}

\date{\today}
\begin{abstract}
The potential energy landscape (PEL) of supercooled binary
Lennard-Jones (BLJ) mixtures exhibits local minima, or inherent
structures (IS), which are organized into meta-basins (MB). We study
the particle rearrangements related to transitions between both
successive IS and successive MB for a small 80:20 BLJ system near the
mode-coupling temperature $T_{MCT}$. The analysis includes the
displacements of individual particles, the localization of the
rearrangements and the relevance of string-like motion. We find that
the particle rearrangements during IS and MB transitions do not
change significantly at $T_{MCT}$. In particular, an onset of single
particle hopping on the length scale of the inter-particle distance
is not observed. Further, it is demonstrated that IS and MB dynamics
are spatially heterogeneous and facilitated by string-like motion. To
investigate the mechanism of string-like motion, we follow the
particle rearrangements during suitable sequences of IS transitions.
We find that most strings observed after a series of transitions do
not move coherently during a single transition, but subunits of
different sizes are active at different times. Several findings
suggest that, though string-like motion is of comparable relevance
when the system explores a MB and when it moves from one MB to
another, the occurrence of a \emph{successful} string enables the
system to exit a MB. Moreover, we show that the particle
rearrangements during two consecutive MB transitions are basically
uncorrelated. Specifically, different groups of particles are highly
mobile during subsequent MB transitions. We further find the positions of 
strings during successive MB transitions weakly but positively correlated 
supporting the idea of dynamic facilitation. Finally, the relation 
between the features of the PEL and the relaxation processes in supercooled 
liquids is discussed.

\end{abstract}
\maketitle

\section{Introduction}
For supercooled liquids approaching the glass transition temperature
$T_g$, the viscosity increases continuously by more than 13 orders of
magnitude. Despite valuable progress in recent years, a complete
theory that rationalizes the corresponding slowing down in molecular
dynamics is still lacking. On the one hand, mode coupling theory
(MCT) predicts a power-law divergence of relaxation times at a
critical temperature $T_{MCT}$~\cite{Goetze}. Though such a
temperature dependence provides a satisfactory description of
experimental data above $T_{MCT}$~\cite{Goetze2}, a singularity at
the critical temperature is not observed and $T_{MCT}$ is typically
$1.1\!-\!1.5\,T_g$. This deviation is usually attributed to hopping
processes that restore ergodicity below $T_{MCT}$, but are not
included in the ideal MCT. On the other hand, potential energy
landscape (PEL) approaches, based on the pioneering work of Goldstein
\cite{Goldstein}, have proven useful in the field of the glass
transition phenomenon~\cite{Angell,Stillinger,Debenedetti}. Here, the
high-dimensional vector of all particle coordinates is considered as
a point moving on the potential energy surface. A PEL description is
particularly useful at sufficiently low temperatures where, as
proposed by Goldstein~\cite{Goldstein}, the evolution of the system
can be decomposed into vibrations about local minima of the PEL on
short timescales and transitions among distinct minima on long
timescales. Then, knowing the properties of the local minima of the
PEL is often sufficient to calculate observables.

In computational work on supercooled liquids, a detailed picture of
the PEL is accessible. Following Goldstein's idea, Stillinger and
Weber~\cite{Stillinger-Weber} introduced the concept of basins where
a basin in configuration space is defined as the set of points that
via steepest descent path along the PEL maps onto the same local
minimum, or inherent structure (IS). The resulting tiling of the
configuration space into non-overlapping basins of attraction
simplifies the thermodynamic description of the system. For example,
it is possible to write the free energy approximately as a function
of the energies and the vibrational frequencies of the IS
\cite{Sciortino,Buechner} so that the statistical properties of these
minima determine all thermodynamic quantities. A link between
thermodynamics and kinetics is suggested by the Adam-Gibbs relation
which connects the relaxation times with the configurational entropy
\cite{AG}. Such a link was qualitatively established by Sastry et
al.~\cite{Sastry-Debenedetti} who observed that the onset of
non-exponential relaxation is accompanied by the sampling of IS with
progressively lower potential energy upon cooling. Later, the
Adam-Gibbs equation was shown to hold quantitatively in simulations
of water~\cite{Scala}, a binary Lennard-Jones (BLJ)
liquid~\cite{Sastry} and silica~\cite{Saika-Voivod}. Further, a
dependence of the fragility on the shape of the PEL was reported
\cite{Sastry,Saika-Voivod}. Such a relation was long proposed by
Stillinger who suggested that strong liquids have a uniformly rough
PEL, while the IS of fragile liquids are organized into meta-basins
(MB)~\cite{Stillinger,Debenedetti} or, equivalently, funnels
\cite{Wolynes,Wales}.

Several workers studied the relation between the properties of the
PEL and the trajectories of the individual particles in more detail.
Schr\o der et al.~\cite{Schroder} confirmed Goldstein's picture
\cite{Goldstein} showing that the dynamics of a BLJ liquid below
$T\!\approx\!T_{MCT}$ can be separated into local vibrations and
transitions between IS. Further, string-like displacements of groups
of particles were found during IS transitions similar to the motion
observed in the equilibrium liquid~\cite{Strings}. Finally,
single-particle hopping indicated by a secondary peak in the van Hove
correlation function at $T\!\leq\!T_{MCT}$ was reported not to result
from the crossing of single energy barriers, but from sequences of IS
transitions~\cite{Schroder}. Instantaneous normal mode analysis
suggests that a transition from non-activated to activated dynamics
takes place at
$T_{MCT}$~\cite{Donati,Broderix,Angelani,LaNave,LaNave-2}. At
$T\!>\!T_{MCT}$, the system is always close to a multi-dimensional
ridge between two basins, and the slowing down of dynamics arises
from the reduction of directions along which free exploration of the
configuration space is possible. In contrast, at $T\!<\!T_{MCT}$, the
system samples regions of the configuration space that have no free
directions so that activated processes are required for relaxation.
However, the method to determine a correct distribution of saddle
points is still a matter of debate~\cite{Doye,Heuer-2}. In different
approaches, no indication for a change of the dynamical behavior at
$T_{MCT}$ was observed, but activated barrier crossing was found to
be relevant already above $T_{MCT}$~\cite{Heuer-2,Heuer-1,Reichman}.
Doliwa and Heuer~\cite{Heuer-2,Heuer-1} calculated the diffusion
constant $D(T)$ from the static properties of the PEL by taking into
account the organization of the IS into MB. In this way, the effect
of back-and-forth jumps, which frequently occur between the IS within
the same MB, but do not contribute to long-range diffusion, can be
eliminated, and only hopping between MB is considered, which should
resemble a random walk on the PEL. It was further shown that a
sequence of jumps is required to escape from low-energy MB. According
to Denny et al.~\cite{Reichman} ``transitions between IS within a MB
involve very small flexing of a cage, while transitions between MB
involve collective rearrangements''. Wales and coworkers
distinguished between ``non-diffusive'' and ``diffusive''
rearrangements depending on whether nearest neighbors are changed and
found that the energy barriers involved in both types of
rearrangements differ more in strong than in fragile liquids
\cite{Middleton}. Further, they observed that upon cooling the
distance atoms move during an IS transition decreases significantly
at $T\!\approx\!T_{MCT}$~\cite{Rojas}.

Despite this progress, the relation between the dynamics of single
particles in real space and the motion of the system in configuration
space is still elusive. In particular, the gap separating MCT and PEL
approaches has not yet been bridged~\cite{Debenedetti}. In this
article, we present a detailed study of the particle rearrangements
resulting from transitions between local minima of the PEL of a
supercooled BLJ liquid. Due to the organization of IS into MB, we
consider transitions between consecutive IS and consecutive MB where
the latter result from a series of the former. For simplicity, we
refer to the corresponding displacements as IS dynamics and MB
dynamics, respectively. Stillinger~\cite{Stillinger} related MB
dynamics to the $\alpha$-process and IS dynamics to the
(Johari-Goldstein) $\beta$-process~\cite{JG} and, hence, our approach
may yield valuable insights into the nature of these relaxation
processes in supercooled liquids. The present analysis of IS/MB
dynamics includes measures for the displacements of the individual
particles, the localization of the rearrangements and the relevance
of string-like motion, which is believed to facilitate the structural
relaxation in supercooled liquids~\cite{Strings,Glotzer,Schober}. We
focus on two temperatures $T_h\!\!>\!T_{MCT}$ and
$T_l\!\leq\!T_{MCT}$. In this way, we can investigate whether a
change of the transport mechanism observed for the equilibrium liquid
at
$T\!\approx\!T_{MCT}$~\cite{Sastry-Debenedetti,Schroder,Roux,Wahnstrom},
cf.\ section \ref{model}, manifests itself in distinct IS/MB
dynamics. Further, by following the particle displacements during
suitably defined sequences of IS transitions, we study the mechanism
of string-like motion in detail. To analyze IS/MB dynamics beyond the
concept of single transitions, we investigate the correlation of the
particle displacements during consecutive IS/MB transitions. In view
of our results and recent experimental findings, we suggest a
modification of Stillinger's picture of the relation between the
features of the PEL and the relaxation processes in supercooled
liquids.

\section{Model and Simulation}\label{model}

The simulations used to generate the data analyzed here are described
in Ref.~\cite{Heuer-2}. In summary, we investigate a BLJ liquid
characterized by the interaction potential
\begin{equation}
V_{\alpha\beta}(r)=4\varepsilon_{\alpha\beta}\,[(\sigma_{\alpha\beta}/r)^{12}
-(\sigma_{\alpha\beta}/r)^{6}]
\end{equation}
with the parameters $N\!=\!N_A\!+\!N_B\!=\!65$, $N_A\!=\!52$,
$\sigma_{AB}\!=\!0.8\,\sigma_{AA}$,
$\sigma_{BB}\!=\!0.88\,\sigma_{AA}$,
$\varepsilon_{AB}\!=\!1.5\,\varepsilon_{AA}$,
$\varepsilon_{BB}\!=\!0.5\,\varepsilon_{AA}$ and $r_c\!=\!1.8$.
Linear functions are added to the potentials to ensure continuous
forces and energies at the cutoff $r_c$. These modifications of the
original potential of Kob and Andersen~\cite{Kob,Kob-2} are necessary
for the simulation of small systems~\cite{Broderix,Rojas}. The data
were generated using Langevin molecular dynamics simulations (MD)
with fixed step size $\lambda\!=\!0.015\!=\!(2k_BT\Delta
t/m\varsigma)^{1/2}$, equal particle masses $m$, friction constant
$\varsigma\!=\!1$ and periodic boundary conditions. All results are
reported in units of $\sigma_{AA}$, $m$, $\varepsilon_{AA}$ and
$m\varsigma\lambda^2/2\varepsilon_{AA}$ for length, mass, energy and
time, respectively.

At regularly spaced times $t_i$, we quench the MD trajectory $x(t)$
to the bottom of the visited basins~\cite{Heuer-2}, yielding the
discontinuous trajectory of IS, $\xi(t_i)$~\cite{Stillinger-Weber}.
In the IS trajectory, the entries at subsequent $t_i$ are identical
until a transition to a new basin of attraction takes place. These
multiple entries are eliminated so that the resulting trajectory only
contains the IS separated by transitions. Further, we apply an
interval bisection method described in Ref.~\cite{Heuer-2} to ensure
that all relevant transitions between the regular quenches are
resolved. The final set of trajectories $\xi(t_j)$ contains all
successively visited IS and consists of more than $10^4$ IS for the
studied temperatures $T_l\!=\!0.435$ and $T_h\!=\!0.500$. From
$\xi(t_j)$, we extract the $N$ trajectories of the individual
particles, $\vec{r}^{\,i}(t_j)$, which form the basis of the present
analysis. In addition, a straightforward algorithm is applied to
construct the MB~\cite{Heuer-2,Buechner-2}. First, all time intervals
$[t_j^f,t_j^l]$ are searched where $t_j^f$ is the time of the first
and $t_j^l\!\neq\!t_j^f$ the time of the last occurrence of the IS
$\xi(t_j)$. Next, any two overlapping time intervals
$[t_j^f,t_j^l]\cup[t_k^f,t_k^l]$ are combined. A similar procedure
was used in Ref.~\cite{Heuer-2} to calculate the diffusion constant
$D(T)$ from the properties of the constructed MB. Strictly speaking,
IS that occur exactly once in the trajectory $\xi(t_j)$
($t_j^f\!=\!t_j^l$) form trivial MB. Here, we exclusively use the
term MB for non-trivial MB. Thus, the MB transitions studied take
place between non-trivial MB and consist of several IS transitions.

For PEL approaches, it is essential to use small systems, because
otherwise interesting effects are averaged out
\cite{Buechner-2,Keyes,Grigera}. On the other hand, the system should
not be too small so as to avoid significant finite size effects. As
was discussed previously~\cite{Heuer-2,Heuer-3}, $N\!=\!65$ turns out
to be a good compromise. Specifically, comparing the dynamics of
systems with $N\!=\!65$, $N\!=\!135$ and $N\!=\!1000$ it was observed
that the respective diffusion constants $D(T)$ are identical within
20\% for all temperatures $T\!\geq\!T_{MCT}$~\cite{Heuer-3}. Further,
a critical temperature $T_{MCT}\!=\!0.45\!\pm\!0.01$ is obtained from
a power-law fit to $D(T)$ for $N\!=\!65$ as compared to a value
$T_{MCT}\!=\!0.435$ established for larger BLJ
systems~\cite{Kob,Kob-2}. Finally, the behavior of the $N\!=\!130$
system resembles that of two independent copies of the $N\!=\!65$
system~\cite{Heuer-3}. Therefore, significant finite size effects are
absent for $T\!\approx\!T_{MCT}$, and -- similar to what is known for
larger BLJ systems~\cite{Kob,Kob-2} -- the $N\!=\!65$ BLJ system can
be regarded as a model of a typical supercooled liquid.

In previous work on the studied system~\cite{Heuer-2,Heuer-1}, Doliwa
and Heuer calculated the mean waiting time in the MB,
$<\!\tau_{MB}(T)\!>$. Its relation to the time constant of the
$\alpha$-process, $\tau_{\alpha}$, depends on the system size. For
$N\!=\!65$, a ratio
$\tau_{\alpha}/\!\!<\!\!\tau_{MB}\!\!>\,\approx\!30$ is found at
$T_l$ and $T_h$. Further, the mean waiting times in the IS,
$<\!\!\tau_{IS}(T)\!\!>$, are approximately a factor of six shorter
than the corresponding $<\!\!\tau_{MB}(T)\!\!>$. For the diffusion
constants, a ratio $D(T_h)/D(T_l)\!\approx\!10$ was reported
~\cite{Heuer-2,Heuer-1}. However, for various models of equilibrium
liquids, the transport mechanism changes significantly in a
comparable temperature range. Specifically, a secondary peak develops
in the van Hove correlation function upon cooling
$T\!\rightarrow\!T_{MCT}$, which is usually interpreted as an onset
of single-particle
hopping~\cite{Sastry-Debenedetti,Schroder,Roux,Wahnstrom}. Moreover,
$T_h$ lies well in the temperature regime where the idealized version
of the MCT is valid~\cite{Kob,Kob-2}, whereas a strong contribution
of activated processes proposed by the extended version of this
theory can be expected at $T_l\!\leq\!T_{MCT}$. Therefore, despite a
moderate variation of the diffusion constant, the nature of the
structural relaxation may change significantly in the studied
temperature range.

\section{Results}

\subsection{Transitions between inherent
structures}\label{single}

\begin{figure}
\includegraphics[angle=270,width=8.5cm]{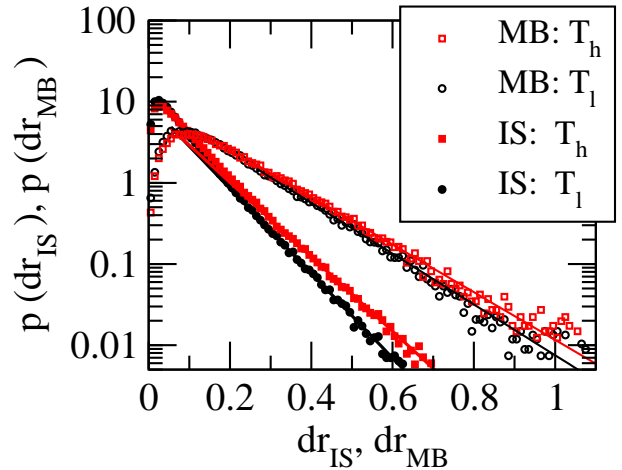}
\caption{
Probability distributions $p\,(dr_{IS})$ and
$p\,(dr_{MB})$ characterizing the displacements of the A particles
during transitions between consecutive inherent structures and
consecutive meta-basins, respectively. Data for $T_l\!=\!0.435$ and
$T_h\!=\!0.500$ are compared. Lines: Exponential decays obtained by
interpolating the data at $dr_{IS/MB}\!\geq\!0.3$.
}
\end{figure}

First, we study the particle displacements $dr_{IS}$ resulting from
IS transitions, i.e., the displacement of particle $i$ during the
transition $\xi(t_{j})\!\rightarrow\!\xi(t_{j+1})$ is given by
$dr_{IS}\!=\!|\vec{r}^{\,i}(t_{j+1})\!-\!\vec{r}^{\,i}(t_{j})|$. In
Fig.\ 1, the probability distribution $p\,(dr_{IS})$ for the A
particles of the BLJ liquid is displayed. For both temperatures, a
rapid decay dominates the distributions. Specifically, for
sufficiently large displacements, $p\,(dr_{IS})$ decays
exponentially. Thus, the distributions for $T\!>\!T_{MCT}$ and
$T\!\leq\!T_{MCT}$ do not exhibit a striking difference, but the
temperature dependence manifests itself in variations of the mean
particle displacement ($<\!dr_{IS}(T_l)\!\!>\:=\!0.081$,
$<\!dr_{IS}(T_h)\!\!>\:=\!0.093$). These findings are consistent with
those of Schr\o der et al.\ \cite{Schroder} on a 50:50 BLJ mixture
($N\!=\!800$) where $p\,(dr_{IS})$ for $T\!\approx\!T_{MCT}$ was
described by a power-law $dr_{IS}^{-5/2}$ at small $dr_{IS}$ and by
an exponential decay at large $dr_{IS}$. There, it was concluded that
the latter functional form results from particles taking part in a
local event, whereas the former is caused by particles adjusting to
this event. Consistent with this interpretation, the deviations from
an exponential behavior at small $dr_{IS}$ are less pronounced in our
case of a smaller system. In contrast to Schr\o der et
al.~\cite{Schroder}, we -- except for an analysis of string-like
motion -- do not include the B particles, because the dynamics of the
minority component of the 80:20 mixture, though qualitatively
similar, is somewhat faster~\cite{Kob,Kob-2} so that a superposition
of the respective results may disturb the underlying functional forms
of the distribution functions.

\begin{figure}
\includegraphics[angle=270,width=8.5cm]{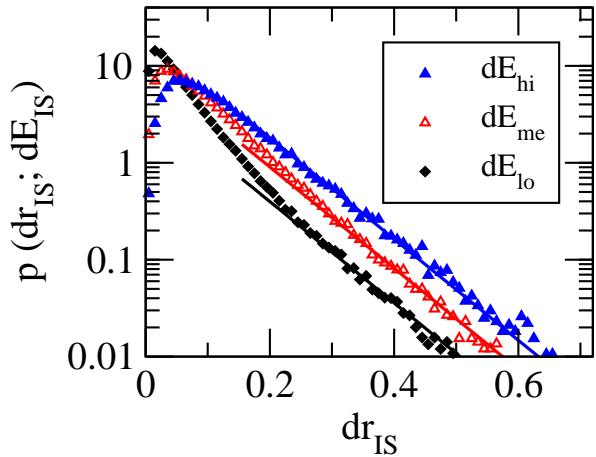}
\caption{
Probability distributions
$p\,(dr_{IS};dE_{IS})$ describing the displacements of the A
particles during transitions between consecutive inherent structures
with energy difference $dE_{IS}$ ($T\!=\!0.435$). We distinguish
between energy differences $dE_{IS}\!<\!0.5$ ($dE_{lo}$),
$0.5\!\leq\!dE_{IS}\!<\!1.5$ ($dE_{me}$) and $1.5\!\leq\!dE_{IS}$
($dE_{hi}$). Lines: Exponential decays $A_i\,\exp(-dr_{IS}/dr_0)$
where the factors $A_i$ were chosen to match the respective curves
$p\,(dr_{IS};dE_{IS})$ and $dr_0$ was obtained from an interpolation
of $p\,(dr_{IS})$, cf.\ Fig.\ 1.
}
\end{figure}

The IS transitions are further characterized by the displacement of
the whole configuration, $dr_{IS}^C\!=\!\sum_{i=1}^{65} dr_{IS}^i$,
and by the absolute value of the energy difference of the involved
IS, $dE_{IS}$. For $T_l$ and $T_h$, we find that the probability
distribution $p\,(dE_{IS})$ is well described by an exponential decay
at sufficiently large $dE_{IS}$. Moreover, the mean energy difference
decreases upon cooling ($<\!\!dE_{IS}(T_l)\!\!>\,=\!0.78$,
$<\!\!dE_{IS}(T_h)\!\!>\,=\!0.98$). The probability distributions
$p\,(dr_{IS}^{C})$ for the studied temperatures are broad and
asymmetric. They exhibit a peak at $dr_{IS}^C\!\approx\!4$ and extend
to $dr_{IS}^C\!\approx\!17$, indicating that the IS dynamics shows a
large diversity. Thus, an analysis in terms of statistical quantities
is necessary.

\begin{figure}
\includegraphics[angle=270,width=8.5cm]{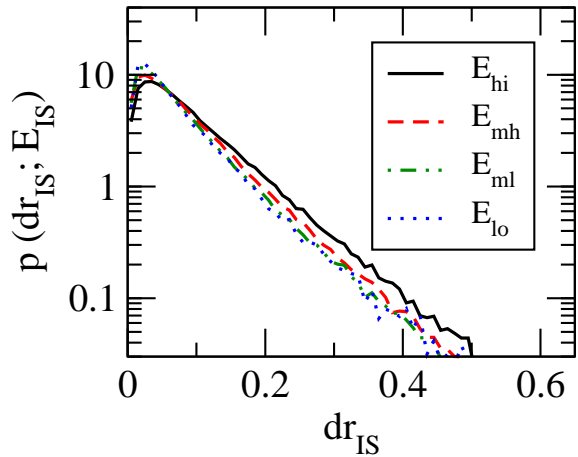}
\caption{
Probability distributions
$p\,(dr_{IS};E_{IS})$ characterizing the displacement of the A
particles during transitions starting from an inherent structure of
energy $E_{IS}$ at $T\!=\!0.435$. We differentiate between energies
$E_{IS}\!<\!-297$ ($E_{lo}$), $-297\!\leq \!E_{IS}\!<\!-295$
($E_{ml}$), $-295\!\leq \!E_{IS}\!<\!-293$ ($E_{mh}$) and $-293\!\leq
\!E_{IS}$ ($E_{hi}$).
}
\end{figure}

Next, the relation between the IS dynamics and the energy difference
$dE_{IS}$ is analyzed by means of the probability distributions
$p\,(dr_{IS};dE_{IS})$. In Fig.\ 2, we see that the mean displacement
of the A particles increases with $dE_{IS}$. For $T_l$,
$<\!dr_{IS}\!\!>\:=\!0.06$ and $<\!dr_{IS}\!\!>\:=\!0.13$ are
obtained for energy differences $dE_{IS}\!<\!0.5$ ($dE_{\,lo}$)
and $dE\!\geq\!1.5$ ($dE_{\,hi}$), respectively. Thus, a large
variation of the potential energy is accompanied by large particle
rearrangements. To study IS dynamics in different regions of the PEL
separately, we consider the probability distributions
$p\,(dr_{IS};E_{IS})$ where $E_{IS}$ is the energy of the initial IS
of the transition. The results for $T_l$ in Fig.\ 3 show that the
energy $E_{IS}$ weakly affects the particle displacements. Hence, IS
dynamics in different regions of the PEL are comparable. We note that
the results for $T_h$ are qualitatively similar to those shown in
Figs.\ 2 and 3. Further, it was checked that the basic features of
our findings do not depend on the specific choice of the energy
limits.

\begin{figure}
\includegraphics[angle=270,width=8.5cm]{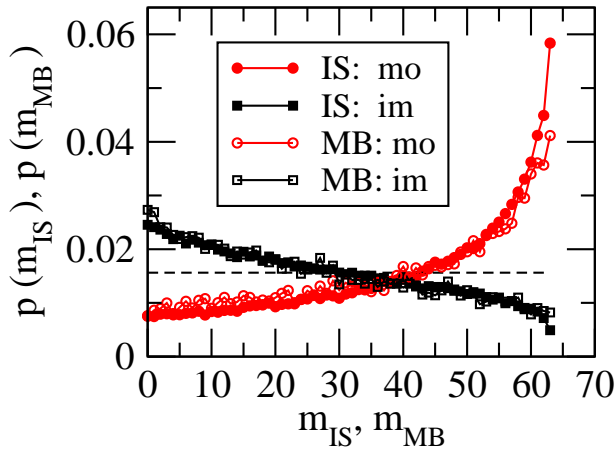}
\caption{
Probability distributions
$p\,(m_{IS})$ and $p\,(m_{MB})$ for the mobilities of the particles
in the next neighbor shells of the A particles with the largest (mo)
and the smallest (im) displacement during the respective inherent
structure and meta-basin transition. See text for details. Dashed
line: $p\,(m_{IS/MB})\!=\!1/64$.
}
\end{figure}

It is known from both experiments~\cite{Heuer-4D,Ediger-4D,Sillescu}
and simulations~\cite{Doliwa,Glotzer,Yeshi,Schober} that molecular
dynamics in supercooled liquids are spatially heterogeneous. We
analyze whether particles experiencing similar displacements due to
IS transitions are spatially correlated, too, as suggested by Schr\o
der et al.~\cite{Schroder}. Since a cluster analysis is limited by
the system size $N\!=\!65$, we focus on the A particles with the
largest and the smallest $dr_{IS}$ during the respective transition
and analyze the mobility of their neighbors. Specifically, we first
rank all particles according to their $dr_{IS}$ and assign mobilities
$m_{IS}$ so that $m_{IS}\!=\!65$ is attributed to the most mobile and
$m_{IS}\!=\!1$ to the most immobile particle. Then, we pick the most
mobile A particle, rearrange where required the mobilities of the
remaining particles ($m_{IS}\!=\!1,\dots,64$) and calculate the
probability $p\,(m_{IS})$ that a particle in the first neighbor shell
of the selected particle has the mobility $m_{IS}$. Finally, this
procedure is repeated for the most immobile A particle. Hence, if
mobile and immobile particles were randomly distributed throughout
the system a probability distribution $p\,(m_{IS})\!=\!1/64$ would
result for both the most mobile and the most immobile A particle.
Fig.\ 4 shows that a random distribution does not apply. Instead, the
particles with the largest displacements are mostly surrounded by
other highly mobile particles. Thus, spatially heterogeneous dynamics
is also observed when analyzing IS dynamics of a 65-particle BLJ
liquid close to $T_{MCT}$.

\begin{figure}
\includegraphics[angle=270,width=8.5cm]{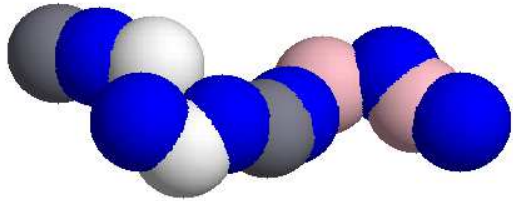}
\caption{
Typical example of string-like
motion during a sequence of transitions between inherent structures.
The gray and the black spheres visualize the particles before and
after the sequence, respectively. The different shades of gray denote
particles replacing their neighbors at different times. It is seen
that the ``macro-string'' after the series contains two
``micro-strings'' due to individual transitions. The micro-string on
the left hand side is active first, the one on the right hand side 5
inherent structure transitions later.
}
\end{figure}

For supercooled liquids close to $T_{MCT}$, it has been demonstrated
that the relaxation of highly mobile particles is facilitated by
''string-like motion'' which means that groups of particles follow
each other in quasi-one-dimensional paths
\cite{Strings,Glotzer,Schober,Aichele,Yeshi2}, see Fig.\ 5. In
addition, such strings were reported for the IS dynamics of a 50:50
BLJ liquid~\cite{Schroder}. Here, we quantify the contribution of
this dynamical pattern to the IS dynamics in more detail. Similar to
Donati et al.~\cite{Strings}, we construct strings by connecting any
two particles $i$ and $k$ if
$\mathrm{min}[\,|\vec{r}^{\,i}(t_j)\!-\!\vec{r}^{\,k}(t_{j+1})|,|\vec{r}^{\,i}(t_{j+1})\!-\!\vec{r}^{\,k}(t_{j})|\,]
\!<\!\delta\!=\!0.6$, where we now consider both the A and the B
particles of the mixture. Since $\delta$ is smaller than the
hard-core radii of the A and the B particles, this condition implies
that one particle has moved and another particle has occupied its
position. With this definition, string-like motion is observed during
29\% of the transitions at $T_l$. On the other hand, for 86\% of
those transitions where string-like motion occurs, the most mobile
particle is involved in a string. This is consistent with a spatial
correlation of highly mobile particles, cf.\ Fig.\ 4.

The strings can be further characterized by their length $l$, i.e.,
the number of participating particles. In Fig.\ 6, we display the
mean number of strings of length $l$ involved in one transition,
$<\!n_{IS}(l)\!\!>$. This quantity is related with the average number
of particles moving in strings during one IS transition by
$<\!\mathcal{N}_{IS}\!\!>\:=\!\sum_l l \!\!<\!n_{IS}(l)\!\!>$. For
$T_l$ and $T_h$, the decrease of $<\!n_{IS}(l)\!\!>$ is consistent
with an exponential decay. However, we find that the functional form
of the decay depends somewhat on the choice of $\delta$. Moreover,
finite size effects can be expected for $l\!\geq\!5$ due to the small
system size. In any case, exponential decays were also observed for
the probability distribution of the string length when analyzing the
dynamics of equilibrium liquids
\cite{Strings,Glotzer,Aichele,Yeshi2}. We emphasize that the observed
temperature dependence of $<\!\mathcal{N}_{IS}\!\!>$ does not imply
that string-like motion is more important at higher temperatures.
Instead, one must take into account that only particles that show a
certain minimum $dr_{IS}$ can fulfill the above specified replacement
criterion and, thus, a larger mean displacement -- as was observed at
the higher temperature, cf.\ Fig.\ 1 -- tends to lead to a larger
number $<\!\mathcal{N}_{IS}\!\!>$. Indeed, it was found for
equilibrium liquids that the fraction of particles moving in strings
increases upon cooling~\cite{Strings,Glotzer}.

\begin{figure}
\includegraphics[angle=270,width=8.5cm]{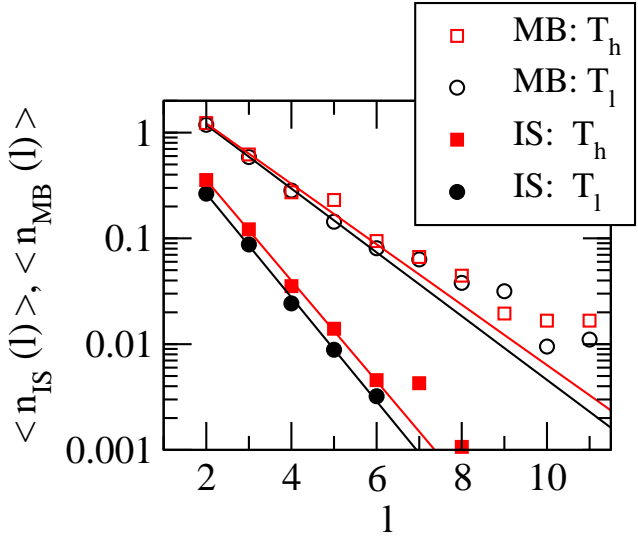}
\caption{
Mean number of strings of length
$l$ observed during transitions between successive inherent
structures ($<\!n_{IS}(l)\!\!>$) and meta-basins
($<\!n_{MB}(l)\!\!>$), respectively. Data for temperatures
$T_l\!=\!0.435$ and $T_h\!=\!0.500$ are shown. Lines: Interpolations
with an exponential decay.
}
\end{figure}

Finally, we study the localization of particle rearrangements
resulting from IS transitions. For this purpose, we measure the
number of involved particles by the quantities $z_{1,IS}$ and
$z_{2,IS}$. The former is defined as
\begin{equation}\label{z1}
z_{1,\,IS}=\sum_{i\varepsilon A}\frac{dr^i_{IS}}{dR_{IS}}
\end{equation}
where $dR_{IS}$ is the maximum displacement of an A particle during
the respective transition, and the latter is calculated according to
\cite{Middleton,Stillinger-Weber-2}
\begin{equation}\label{z2}
z_{2,\,IS}=\frac{\left[\,\sum_{i\varepsilon A}
\,(dr^i_{IS})^2\,\right]\,^2} {\sum_{i\varepsilon
A}\,(dr^i_{IS})^4}\,.
\end{equation}
In the case that $n$ particles move the same distance and the
remainder is immobile, $z_{1,IS}$ and $z_{2,IS}$ equal $n$. In Fig.\
7, we show the probability distribution $p\,(z_{1,IS})$ for the
studied temperatures. For $T_l$ and $T_h$, the distributions are
nearly symmetric and peak at $z_{1,IS}\!\approx\!16$. The shape of
all curves can be described by a Gaussian with a width parameter
$\sigma\!\approx\!4.5$, which again reflects the diversity of IS
dynamics. The distributions $p\,(z_{2,IS})$ (not shown) are close to
a Gaussian centered at $z_{2,IS}\!\approx\!17$ and characterized by
$\sigma\!\approx\!7.5$. A closer inspection reveals that the mean
values $<\!\!z_{1,IS}\!\!>$ and $<\!\!z_{2,IS}\!\!>$ decrease by
about 0.5 when decreasing the temperature from $T_h$ to $T_l$, i.e.,
IS dynamics becomes slightly more local upon cooling. We emphasize
that the value of about 16 for the number of particles involved in IS
dynamics should not be taken too literally since measures of the
localization different from $z_{1,IS}$ and $z_{2,IS}$ can yield
values that are up to a factor of 3 smaller.
\begin{figure}
\includegraphics[angle=270,width=8.5cm]{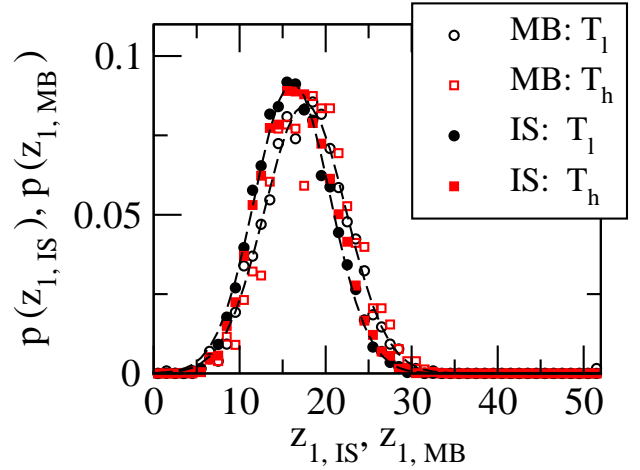}
\caption{
Probability distributions
$p\,(z_{1,IS})$ and $p\,(z_{1,MB})$ characterizing the number of
particles participating in the rearrangements due to transitions
between successive inherent structures and
meta-basins, respectively, cf.\ Eq.\ \ref{z1}. The results for
$T_l\!=\!0.435$ and $T_h\!=\!0.500$ are compared. Dashed lines:
Interpolations of the data for $T_l$ with a Gaussian.
}
\end{figure}

\subsection{Transitions between meta-basins}

Recalling that Stillinger~\cite{Stillinger,Debenedetti} related the
$\alpha$-process to MB transitions, we now turn to the particle
rearrangements during these transitions. For this purpose, we search
the IS with the lowest energy in each MB, $\xi(t_k)$, where $k$ is
the index of the MB, and define the particle displacement during a MB
transition as the one obtained by a comparison of $\xi(t_k)$ and
$\xi(t_{k+1})$. Since $k$ and $k\!+\!1$ are not the indices of
consecutive IS, but of consecutive MB, these displacements are caused
by a sequence of IS transitions. In what follows, we describe the
particle displacement during a MB transition by the vector
$\vec{dr}_{MB}(k)$ and denote its absolute value as $dr_{MB}$.

The probability distributions $p\,(dr_{MB})$ calculated for $T_l$ and
$T_h$ are included in Fig.\ 1. Since MB dynamics results from several
IS transitions, the mean displacement $<\!\!dr_{MB}(T)\!\!>$ is
larger than $<\!\!dr_{IS}(T)\!\!>$. For the temperature dependence of
MB dynamics, we find different mean values
$<\!dr_{MB}(T_l)\!\!>\:=\!0.18$ and $<\!dr_{MB}(T_h)\!\!>\:=\!0.20$,
but the shape of the distributions $p\,(dr_{MB})$ at $T_l$ and $T_h$
is comparable. Similar to $p\,(dr_{IS})$, the distributions are well
described by an exponential decay at large $dr_{MB}$. Interestingly,
for neither type of transitions, there is enhanced probability at
$dr_{IS/MB}\!\approx\!1.0$. Hence, single-particle hopping on the
length scale of the inter-particle distance observed for various
larger models of equilibrium liquids at
$T\!\approx\!T_{MCT}$~\cite{Sastry-Debenedetti,Schroder,Roux,Wahnstrom}
does not manifest itself in IS and MB dynamics of our small system.
The mean energy difference of the IS involved in the MB transition,
$<\!\!dE_{MB}\!\!>$, increases from 1.42 at $T_l$ to 1.56 at $T_h$.

To analyze whether particles showing similar mobilities during MB
transitions are spatially correlated we calculate the probability
distributions $p\,(m_{MB})$ which, analogous to $p\,(m_{IS})$,
characterize the mobility of the neighbors of the most mobile and the
most immobile A particle during a MB transition. It is evident from
Fig.\ 4 that the most mobile particle during a MB transition is
prevailingly surrounded by other mobile particles. Moreover, a
comparison of $p\,(m_{IS})$ and $p\,(m_{MB})$ shows that the
spatially heterogeneous nature of IS and MB dynamics is very similar
at $T\!\approx\!T_{MCT}$. To gain further valuable insights, we study
the localization of the particle rearrangements due to MB transitions
using the quantity $z_{1,MB}$ defined in analogy to Eq.\ \ref{z1}.
Inspecting the probability distributions $p\,(z_{1,MB})$ for $T_l$
and $T_h$ in Fig.\ 7, we see that a comparable number of particles
participates in IS and MB dynamics, suggesting that basically the
same group of particles is ``active'' during all IS transitions
involved in the respective MB transition. Further, similar to IS
dynamics, MB dynamics is slightly less local at the higher
temperature ($<\!\!z_{1,MB}(T_l)\!\!>\,=17.7$,
$<\!\!z_{1,MB}(T_h)\!\!>\,=18.3$). Hence, we find no indication that
the length scales attributed to the spatially heterogeneous nature of
IS and MB dynamics, respectively, increase upon cooling.

We further explore the spatial heterogeneities associated with MB
dynamics by investigating the relevance of string-like motion. With
the same definition as in the previous section, we find that
string-like motion occurs during 87\% of all MB transitions at $T_l$
where the average number of particles that move in strings during a
MB transition amounts to $<\!\mathcal{N}_{MB}\!\!>\:=\!7.2$. These
numbers indicate that string-like motion yields an important
contribution to the particle rearrangements during MB transitions. In
other words, the group of particles that takes part in MB dynamics
usually achieves the large displacements by means of string-like
motion. The relevance of this type of motion for MB dynamics is also
obvious from Fig.\ 6 where we display the distributions
$<\!\!n_{MB}(l)\!\!>$ characterizing the mean number of strings of
length $l$ during one MB transition.

\subsection{Sequences of inherent structure transitions}

\begin{figure}
\includegraphics[angle=270,width=8.5cm]{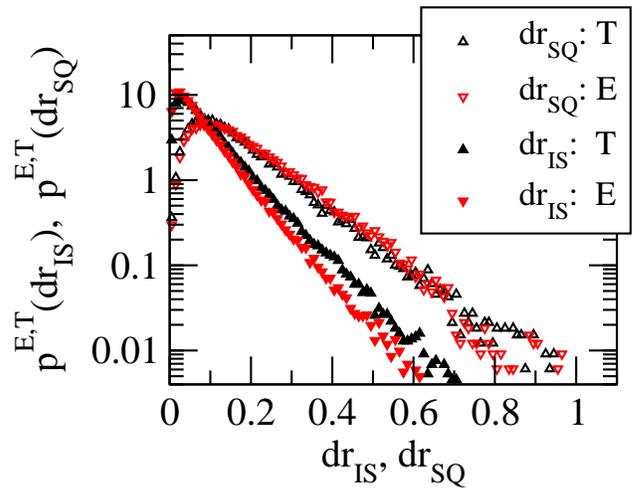}
\caption{
Probability distributions
$p^{\,E}(dr_{SQ})$ and  $p^{\,T}(dr_{SQ})$ describing the displacements
of the A particles during sequences of inherent structure transitions
that occur when a meta-basins is explored (E) and when a transition
between different meta-basins takes place (T), respectively ($T\!=\!0.435$).
See text for details. For comparison, we included the
corresponding results for the involved individual transitions,
$p^{\,E}(dr_{IS})$ and $p^{\,T}(dr_{IS})$.
}
\end{figure}

Due to the organization of the IS into MB, two different processes
are relevant when the system samples the PEL. First, the exploration
of a MB, i.e., the back-and-forth jumps between the IS in the same MB
and, second, the transition between distinct MB. In this section, we
study the particle rearrangements involved in these processes by
following the particle rearrangements during suitable sequences of IS
transitions. Since the diffusion constant is basically determined by
the trapping of the system in long-lived MB~\cite{Heuer-2,Heuer-1},
we now focus on MB within which at least six IS transitions occur,
i.e., we consider 49\% of all MB. To investigate the exploration
process and the transition process separately we construct E- and
T-sequences using the following criteria: An E-sequence combines all
IS transitions connecting the two IS within the same MB that show the
largest distance $dr_{IS}^C$. This means that, on average, nine IS
transitions form an E-sequence for $T_l$. To obtain the T-sequences
we unite four IS transitions that occur when the system moves to
another MB. More precisely, if two successive MB are separated by
more than four IS transitions, we choose four jumps in the middle of
this series. On the other hand, if the MB are separated by less than
four IS transitions, we add the adjoining transitions within the
involved MB to the sequence. We do not regard the latter procedure as
a serious problem because, in any case, long-lived MB are exited by a
series of events~\cite{Heuer-2} so that it is not completely clear
which individual transitions are a part of the escape process. The
chosen numbers are also motivated by the overall displacements during
the E- and T-sequences, see below. However, we ensured that the basic
features of our results are unchanged when the numbers are varied in
a meaningful range. The IS transitions involved in the E- and
T-sequences, respectively, are denoted as E- and T-transitions in
what follows.

Fig.\ 8 shows the probability distributions $p^{\,E,T}(dr_{SQ})$ and
$p^{\,E,T}(dr_{IS})$ for the displacements of the A particles during
the E/T-sequences and the E/T-transitions at $T_l$, respectively.
While different mean displacements
$<\!dr_{IS}\!\!>^{\,T}>\:<\!dr_{IS}\!\!>^{\,E}$ are found for the
single transitions, the rearrangements during the E- and the
T-sequences are described by nearly identical distribution functions.
Thus, the smaller mean displacements during the E--transitions and
the back-and-forth jumps between IS within the same MB lead to the
effect that four T-transitions result in the same overall
displacement as, on average, nine E-transitions. In other words, the
particle rearrangements during the exploration and the transition
process are very similar at the level of sequences of transitions.
Despite the quantitative differences, the distributions for the
single transitions, $p^{\,E}(dr_{IS})$ and $p^{\,T}(dr_{IS})$ are
still comparable. In particular, due to the correlation between the
particle displacement and the energy difference $dE_{IS}$, cf.\ Fig.\
2, some of the deviations result from the fact that larger mean
energy differences are found for the T-transitions. Strictly
speaking, when calculating the distributions
$p^{\,E}(dr_{IS};dE_{IS})$ and $p^{\,T}(dr_{IS};dE_{IS})$ we find
that both distributions are very similar for IS transitions
characterized by high energy differences, but somewhat different in
the case of small $dE_{IS}$.

\begin{figure}
\includegraphics[angle=270,width=8.5cm]{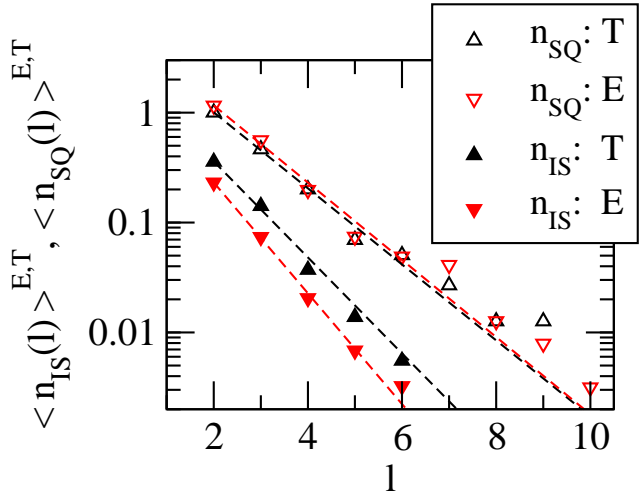}
\caption{
Mean number of strings of length
$l$ resulting from: (i) single transitions between successive
inherent structures visited during the exploration ($<\!n_{IS}(l)\!\!>^{\,E}$)
and the transition process ($<\!n_{IS}(l)\!\!>^{\,T}$), (ii) sequences
of transitions of inherent structures visited during the exploration
($<\!n_{SQ}(l)\!\!>^{\,E}$) and the transition process
($<\!n_{SQ}(l)\!\!>^{\,T}$) ($T\!=\!0.435$). See text for details.
Lines: Interpolations with an exponential decay.
}
\end{figure}

The finding $p^{\,E}(dr_{SQ})\!\approx\!p^{\,T}(dr_{SQ})$ enables us
to study the relevance of string-like motion for the exploration and
the transition process without any effect of the respective mean
particle displacement, see the discussion of Fig.\ 6. This was one
reason for the specific choice of the parameters when constructing
the sequences. The distributions $<\!n_{SQ}(l)\!\!>^{\,E,T}$ for the
mean number of strings of length $l$ during one E- and one T-sequence
at $T_l$, respectively, are displayed in Fig.\ 9. The good agreement
indicates that string-like motion is of similar importance for the
particle rearrangements during the exploration and the transition
process, respectively. This finding excludes that it is simply the
occurrence of string-like motion that allows the system to escape
from a MB. This point will be further discussed later in this
section. For comparison, we include the distributions
$<\!n_{IS}(l)\!\!>^{\,E,T}$ characterizing string-like motion during
the individual E- and T-transitions, in Fig.\ 9. Both distributions
can be interpolated by an exponential decay, suggesting that the
basic features of string-like motion are comparable for the different
types of transitions. As was discussed above, the quantitative
differences result at least in part from the distinct mean particle
displacements due to the E- and the T-transitions, cf.\ Fig.\ 8.
Taking also into account that a comparable number of particle
participates in E- and T-transitions
($<\!\!z_{1,IS}\!\!>^{\,E}\,=\!16.3$,
$<\!\!z_{1,IS}\!\!>^{\,T}\,=\!16.3$ for $T_l$) we conclude that the
different types of transitions show at most some quantitative
differences in the vicinity of $T_{MCT}$, supporting our prior
finding that IS dynamics in different regions of the PEL are
comparable, see Fig.\ 3.

Next, we study the mechanism of string-like motion. When comparing
$<\!n_{SQ}(l)\!\!>^{\,E,T}$ with their counterparts for the single
transitions $<\!n_{IS}(l)\!\!>^{\,E,T}$, cf.\ Fig.\ 9, it becomes
obvious that the distributions have different slopes in a
semi-logarithmic representation. Hence, one may speculate that a
string observed during a sequence does not result from a single
transition, but from the interplay of particle displacements due to
subsequent jumps. For example, one can imagine that a long
``macro-string'', defined as a string during a sequence, is composed
of several ``micro-strings'' resulting from the displacements during
the individual transitions of the series. This will now be studied in
more detail. For this purpose, we consider all macro-strings of
length $l\!\geq\!3$ and analyze how many of them result from a single
transition. We find that only 15.4\% and 13.7\% of the macro-strings
during the E- and the T-sequences, respectively, are the consequence
of a single event. On the other hand, 48.4\% (E) and 57.5\% (T) of
the macro-strings are composed of at least one micro-string. The
remainder, namely, 36.2\% (E) and 28.8\% (T), can be traced back to
concerted single-particle type displacements in successive jumps.
Likewise, for 47.8\% (E) and 53.8\% (T) of all particles involved in
macro-strings of length $l\!\geq\!3$, the replacement of the
neighboring particle takes place during a single transition, while,
for the remainder of the particles, several displacements must add up
for the replacement criterion to be fulfilled. Of course, these
numbers depend on the definition of the strings. Nevertheless, they
show that most macro-strings, especially, the long ones, do not
result from a coherent motion of all particles, but from subsequent
motions of single particles or small groups of particles.

A typical example of the interplay of micro-strings and
single-particle motions in forming a large macro-string is shown in
Fig.\ 5, where the different shades of gray denote particles
replacing each other at different times. It is also seen that the
replacements do not start and end at the ``head'' (right hand site)
and the ``tail'' (left hand side) of the string, respectively, but
occur in a random order. Though ordered sequential replacements along
the string are observed in many instances, we still regard the
scenario in Fig.\ 5 as a typical example.

Especially, the macro-strings of the E-sequences may be a subtle
result of individual motions. Since the IS within a MB are visited
several times, the micro-strings of the E-sequences may show
back-and-forth jumps. To quantify this effect we calculated the
probability that a micro-string observed during an E-transition jumps
back during a later transition in the same MB. We find that for
44.5\% of the micro-strings all particles jump back to their initial
positions. Hence, back-and-forth jumps of micro-strings are a
frequent phenomenon when observing IS dynamics within a MB. In
comparison, the back-jump probability for the micro-strings during
the T-sequences amounts to only 2\%. Considering also the results in
Fig.\ 9, one may speculate that, though string-like motion is of
similar relevance for the particle rearrangements during the
exploration and the transition process, respectively, it is the
occurrence of \emph{successful} strings that enables the system to
escape from a MB.

\subsection{Correlation of successive particle displacements}

As aforementioned, results of Doliwa and Heuer~\cite{Heuer-2,Heuer-1}
suggest that jumps between MB resemble a random-walk on the PEL. To
check this conlusion, we study the correlation of particle
displacements resulting from MB transitions at two different times,
i.e., four-time correlation functions are considered. To measure for
how many MB transitions an A particle remembers the direction of an
initial motion we define the correlation function
\begin{equation} \label{D}
D(\Delta k)=\left<
\frac{\vec{dr}_{MB}(k)}{dr_{MB}(k)}\cdot\frac{\vec{dr}_{MB}(k\!+\!\Delta
k )}{dr_{MB}(k\!+\!\Delta k)}\right>.
\end{equation}
Here, the brackets $<\!\dots\!>$ denote the average over all MB
transitions and all A particles. Due to the properties of the scalar
product, $D(\Delta k)$ will be positive (negative) if, on average,
the displacements of a particle during the MB transitions
$\xi(t_k)\!\rightarrow\!\xi(t_{k+1})$ and $\xi(t_{k+\Delta k}
)\!\rightarrow\!\xi(t_{k+\Delta k+1})$, respectively, have the same
(opposite) direction. Another property of the particle displacement
during a MB transition is the mobility. Different from the definition
used so far, we now characterize the relative mobility of a particle
by
\begin{equation}
\mu(k)=\frac{dr_{MB}(k)-<\!dr_{MB}(k)\!\!>}{<\!dr_{MB}(k)\!\!>}
\end{equation}
where $<\!dr_{MB}(k)\!\!>$ is the mean displacement of an A particle
during the MB transition $\xi(t_k)\!\rightarrow\!\xi(t_{k+1})$.
Hence, the correlation function
\begin{equation}\label{M}
M(\Delta k)=\left<\mu(k)\,\mu(k\!+\!\Delta k)\right>
\end{equation}
relates the relative mobilities during different MB transitions.
To study the correlation of the particle displacements resulting
from different IS transitions, we define the correlation functions
$D(\Delta j)$ and $M(\Delta j)$ in analogy to their counterparts
for the MB transitions.

\begin{figure*}
\includegraphics[width=13cm]{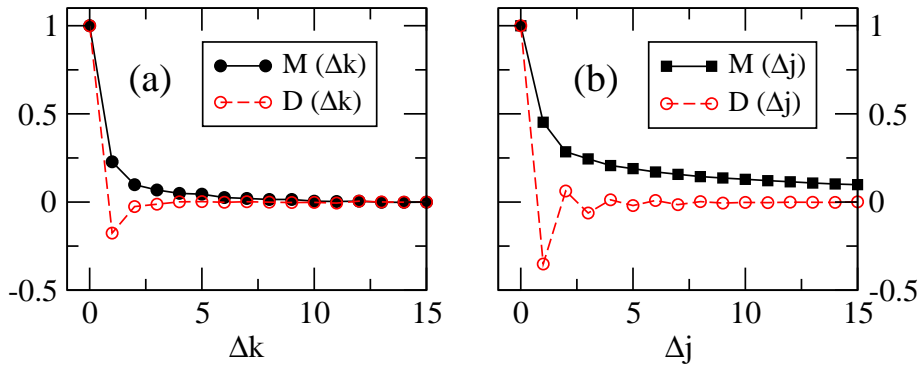}
\caption{
(a): Correlation functions that
relate the displacements of single A particles during different
transitions between meta-basins at $T\!=\!0.435$. $D(\Delta k)$
measures the correlation of the directions of subsequent motions;
$M(\Delta k)$ correlates the respective relative mobilities, cf.\
Eqs.\ \ref{D} and \ref{M}. (b): Analogously defined quantities
$D(\Delta j)$ and $M(\Delta j)$ characterizing the correlation of the
particle displacements during different transitions between inherent
structures.
}
\end{figure*}

Fig.\ 10 (a) shows $D(\Delta k)$ and $M(\Delta k)$ for the MB
dynamics at $T_l$. Inspecting the latter correlation function a rapid
decay is evident. For the former, the negative sign of $D(\Delta
k\!=\!-1)$ implies that subsequent motions are backward correlated,
i.e., the particles tend to move towards their old positions during
the next jump. However, the values $|D(\Delta k\!=\!1)|\!\ll\!1$ and
$M(\Delta k\!=\!1)\!\ll\!1$ indicate that the particle rearrangements
during consecutive MB transitions show only a small correlation.
Hence, a random walk on the PEL is indeed a good approximation for
the jumps between MB. In addition, the findings imply that, during
different MB transitions, different groups of particles are highly mobile and
form strings, corroborating our speculation that the occurrence of a
successful string results in the escape from a MB.

For comparison, $D(\Delta j)$ and $M(\Delta j)$ characterizing IS
dynamics at $T_l$ are displayed in Fig.\ 10(b). Due to the
back-and-forth jumps between the IS within a MB the correlations
persist for several IS transitions. In particular, sequences of
back-and-forth jumps between two IS result in a oscillatory behavior
of $D(\Delta j)$. $M(\Delta j)$ does not completely decay within the
first 15 IS transitions. Since the same transition may occur several
times when the system explores a MB, the information about the
initial mobility is not completely lost until the MB is exited.

Very recently, Garrahan and Chandler~\cite{Chandler,Garrahan}
proposed that bulk dynamics in supercooled liquids can be understood
based on two simple ingredients, namely, the existence of spatially
heterogeneous dynamics and the facilitation of dynamics in the
vicinity of mobile regions. To check this assumption, we study
whether strings during consecutive MB transitions are spatially
correlated. For this purpose, we first characterize the positions of
all strings by their center of mass. Then, for every string during
the MB transition $k\!\rightarrow\!k\!+\!1$, the minimum distance
between its position and the positions of any string during the
transition $k\!-\!1\!\rightarrow\!k$ is determined. Finally, we
repeat this calculation for randomly chosen positions of the strings
during the MB transition $k\!\rightarrow\!k\!+\!1$. For the actual
strings at $T_l$, we obtain a mean minimum distance $<\!\!d\!>\,=1.3$
which is slightly smaller than $<\!\!d\!>\,=1.5$ found for a random
distribution, suggesting that dynamics in the vicinity of mobile
regions may be somewhat facilitated. However, the observed effect is
weak and needs to be validated for larger systems.

\section{Discussion and Summary}

We studied the particle rearrangements during transitions between
consecutive inherent structures (IS) and consecutive meta-basins (MB)
of a supercooled 80:20 BLJ liquid. Specifically, the displacements of
the individual particles, the localization of the rearrangements and
the string-like motion were characterized for two temperatures,
$T_h\!>\!T_{MCT}$ and $T_l\!\leq\!T_{MCT}$. In addition, the IS
dynamics were analyzed in dependence of the energies of the involved
IS and their energy difference. Considering that the motion of the
system on the PEL can be decomposed into the exploration of MB and
the transition between distinct MB, we compared the particle
rearrangements resulting from both processes. This analysis was done
both at the level of single transitions (E- and T-transitions) and at
the level of suitable sequences of transitions (E- and T-sequences).
Finally, time correlation functions of the displacements during
transitions between successive MB were analyzed. Since the time
constants for the dynamics of the A and the B particles of the BLJ
mixture are somewhat different~\cite{Kob,Kob-2}, we mostly focused on
the motion of the former. However, with respect to all studied
quantities, only minor differences are observed for the B particles.

Comparing the results for $T_h\!>\!T_{MCT}$ and $T_l\!\leq\!T_{MCT}$,
no change of the basic features except only gradual variations were
found. For IS and MB dynamics, the mean energy differences,
$<\!dE_{IS/MB}\!\!>$, the mean displacement of the A particles,
$<\!dr_{IS/MB}\!\!>$, and the mean number of A particles taking part
in a transition, $<\!z_{1,IS/MB}\!>$, decrease upon cooling, but the
shape of the corresponding distributions is basically unchanged. For
$T_l$ and $T_h$, the distributions $p\,(z_{1,\,IS/MB})$ are close to
Gaussian and, over a wide range, the curves $p\,(dr_{IS/MB})$ decay
exponentially. The decrease of $<\!dE_{IS/MB}\!\!>$ is consistent
with the decline of $k_BT$. The variation of $<\!dr_{IS/MB}\!\!>$ is,
at least in part, a consequence of the changing mean energy
difference, because it was found that large changes of the energy are
accompanied by large particle displacements. Hence, when observing IS
and MB dynamics in a 65-particle BLJ liquid, there is no evidence
that the mechanism for the particle motion is discontinuously altered
at $T_{MCT}$. In particular, we do not find single-particle hopping
on the length scale of the inter-particle distance.

These results are in agreement with findings by Schr\o der et al.\
\cite{Schroder}, who concluded that single-particle hopping observed
for the equilibrium liquid at $T\!\approx\!T_{MCT}$ does not result
from transitions over \emph{single} barriers, but ``the jump occurs
via a number of intermediate IS''. On the other hand, at variance
with the outcome of the present work, Hernandez-Rojas and
Wales~\cite{Rojas} reported a ``distinct change in behavior'' at
$T\!\approx\!T_{MCT}$ when investigating the particle rearrangements
in a kinetic Monte-Carlo approach. Since $T_l\!=\!0.435$ is still
close to $T_{MCT}\!=\!0.45\!\pm\!0.01$ it may be useful to check our
results over a broader temperature range. With this reservation, the
absence of a significant change of the dynamical behavior at
$T_{MCT}$ appears to support that the sampling of the PEL changes
gradually~\cite{Heuer-2,Reichman} instead of discontinuously as
concluded in normal-mode analysis approaches
\cite{Donati,Broderix,Angelani,LaNave,LaNave-2}.

In addition, we analyzed the dependence of IS dynamics on the energy
of the initial IS, $E_{IS}$. Weak variations of
$<\!dr_{IS}(E_{IS})\!\!>$ indicate that the energy has a minor
influence. These results suggest that IS dynamics in different
regions of the PEL are comparable. Such behavior is in accordance
with the current picture of the PEL of fragile glass formers where a
comparable ruggedness is assumed throughout the
landscape~\cite{Stillinger,Debenedetti}. Moreover, a weak dependence
of IS dynamics on $E_{IS}$ is consistent with our findings for the
particle rearrangements during E- and T-transitions, i.e., for IS
dynamics occurring when a MB is explored and when a transition
between different MB takes place, respectively. We observed that the
mean displacement $<\!dr_{IS}\!\!>^{\,T}$, is somewhat larger than
$<\!dr_{IS}\!\!>^{\,E}$, but this effect results in part from larger
mean energy differences during the T-transitions, because generally
the particle rearrangements are larger for higher energy differences.
In addition, the number of particles involved in IS dynamics and in
string-like motion are similar during the exploration and the
transition process. Hence, with respect to the accompanying particle
rearrangements, no principal differences exist between IS transitions
near the bottom of a MB and near the ridges between distinct MB.

We further demonstrated that the particles participating in IS and MB
dynamics, respectively, i.e., particles showing high displacements
are not randomly distributed, but rather reside prevailingly in the
next neighbor shell of each other. Thus, consistent with previous
computational work on supercooled liquids
\cite{Strings,Doliwa,Glotzer,Yeshi,Schober}, IS and MB dynamics at
$T\!\approx\!T_{MCT}$ are spatially heterogeneous. In
experiments~\cite{Heuer-4D,Ediger-4D}, a spatially heterogeneous
nature was observed for dynamics near $T_g$. Though the dynamical
heterogeneities at $T\!\approx\!T_{MCT}$ and at $T\!\geq\!T_{g}$,
respectively, show similar features, e.g., a comparable rate
memory~\cite{Okun}, their exact relation is still elusive.

The clustering of mobile particles found for IS and MB dynamics is in
accordance with the occurrence of string-like motion. A closer
analysis revealed that this dynamical pattern is observed during 29\%
of the IS transitions and during 87\% of the MB transitions at $T_l$.
The latter value together with $<\!\!\mathcal{N}_{MB}\!\!>\,=\!7.2$
obtained for the mean number of particles forming strings per MB
transition indicate that string-like motion is very important for MB
dynamics. On the other hand, this type of motion appears to be less
relevant on the shorter time scale of IS dynamics. However, one has
to consider that during an IS transition, the displacement of most
particles is too small to satisfy the criterion used to define
strings. Hence, a limited relevance is a natural consequence. To gain
further insights, we compared string-like motion during sequences of
IS transitions that take place when the system explores a MB and when
it moves from one MB to another, respectively. Constructing E- and
T-sequences such that comparable particle displacements result, we
observed $<\!n_{SQ}(l)\!\!>^{\,E}\approx\:<\!n_{SQ}(l)\!\!>^{\,T}$
for the mean number of strings of length $l$ during the respective
sequences. Thus, string-like motion during the exploration and the
transition process is not different, suggesting that the probability
to find a string depends merely on the displacements of the
particles.

Concerning the mechanism of string-like motion, it was demonstrated
that the formation of strings results from the concerted interplay of
the particle rearrangements related to \emph{subsequent} IS
transitions. In other words, most macro-strings defined as strings
that result from a series of transitions do not arise from a coherent
motion of all particles at the same time, but from coordinated
displacements of subunits at different times. These subunits
include both single particles and micro-strings, i.e., small groups
of particles replacing each other within one transition. Further, the
subsequent particle replacements in a macro-string do not necessarily
start at the ``head'' of the string and end at the ``tail'', but
successive replacements often take place at random positions in the
string, cf.\ Fig.\ 5. Moreover, for IS dynamics inside the MB, the
formation of a macro-string frequently includes back-and-forth jumps
of the involved micro-strings. Hence, all these findings elucidate
that string-like motion is a complex dynamical process in which
several rearrangements within a group of particles occur. In view of
this result, the observation
$<\!\!z_{1,IS}\!\!>\,\approx\,<\!\!z_{1,MB}\!\!>$ for the number of
particles taking part in IS and MB dynamics, respectively, becomes
plausible. Since particles that are ``active'' during a MB transition
mostly participate in strings, the multi-step nature of this type of
motion has the consequence that basically the same group of particles
is active during all IS transitions involved in the respective MB
transition, i.e., for several IS transitions, high particle mobility
persists in a certain region of the system. Based on the multi-step
nature of string-like motion, Gebremichael et al.\ \cite{Yeshi2}
speculate that the occurrence of strings is related to local
structural properties of supercooled liquids. For example, they
suggest that temporary fissures open a quasi-1D channel in which the
particles can move in strings. Further investigation on this point is
underway~\cite{Magnus}.

Finally, we studied the correlation of particle displacements during
different transitions between MB, i.e., time correlation functions of
displacements were investigated. Analyzing both the direction of the
motion and the relative mobility we find that the displacements of
any one particular particle during consecutive MB transitions are
basically uncorrelated. In other words, different groups of particles
are mobile during successive MB transitions. In contrast, when
relating the particle rearrangements due to different IS transitions,
the back-and-forth jumps between the IS within a MB and the process
of string formation result in a correlation of the particle
displacements that persists for several IS transitions. We also
investigated whether the strings during two consecutive MB
transitions are spatially correlated. For this purpose, we calculated
the minimum distance between a new string and any of the old strings.
We found a mean minimum distance that is 15\% smaller than in the
case of a random distribution of the new strings, suggesting that new
strings tend to be formed near regions where string-like motion has
taken place. This appears to support Garrahan and
Chandler~\cite{Chandler,Garrahan} who stress the importance of
dynamic facilitation, i.e., they assume that regions showing high
mobility assist neighboring regions to become mobile. However, the
effects observed here are weak and need to be validated for larger
systems, as well as in the equilibrium liquid.

As aforementioned, Stillinger \cite{Stillinger,Debenedetti} assigned
the $\alpha$- and the (Johari-Goldstein) $\beta$-process to MB and IS
dynamics, respectively. From experimental work, it is known that,
different from the $\alpha$-process, the $\beta$-process exhibits an
Arrhenius-like temperature dependence~\cite{JG} so that both
processes often merge into a single high temperature relaxation near
$T_{MCT}$~\cite{Kudlik}. Multi-dimensional NMR experiments have shown
that the molecular reorientations associated with the
$\alpha$-~\cite{Spiess,Roessler,Tracht} and the
$\beta$-process~\cite{Vogel-1,Vogel-2,Vogel-3} close to $T_g$ are
complex multi-step processes. For the latter relaxation, the
multi-step reorientation is restricted to a small section of the unit
sphere~\cite{Hinze,Vogel-1,Vogel-2,Vogel-3} where the accessible
solid angle strongly grows with increasing
temperature~\cite{JG,Kudlik,Vogel-2}. In addition, computational
studies have revealed that the length scale associated with the
spatially heterogeneous nature of the structural relaxation in
supercooled liquids increases upon cooling
$T\!\rightarrow\!T_{MCT}$~\cite{Glotzer,Das,Mountain,Lacevic}.

We now discuss Stillinger's view in the context of the present and
the summarized previous findings for supercooled liquids. Several
results are inconsistent with the assumption that the
$\alpha$-process is identical with MB dynamics. First of all, we
found that MB dynamics results in a mean particle displacement that
is much smaller than the inter-particle distance, cf.\ Fig.\ 1,
indicating that, at the studied temperatures, the structural
relaxation is not complete after a single MB transition. Likewise,
$\tau_{\alpha}$ is approximately a factor of 30 longer than the mean
waiting time in the MB $<\!\!\tau_{MB}\!\!>$~\cite{Heuer-2,Heuer-1}.
Finally, the temperature dependence of the mean number of particles
involved in a MB transition, $<\!z_{1,MB}\!>$, is incompatible with
an increasing length scale attributed to the spatial heterogeneities
of MB dynamics. Concerning the $\beta$-process, some of the
experimental results appear to be at variance with the assumption
that this relaxation is related to single transitions between IS.
Specifically, the experimental finding that the $\beta$-process
results from a multi-step process is inconsistent with a single jump
in the PEL. Moreover, the strong growth of the $\beta$-relaxation
strength is difficult to understand. In particular, an explanation
will be problematic if the IS dynamics in different regions of the
PEL are similar as suggested by our results.

In view of these findings, we propose that the MB transitions are the
elementary steps of the $\alpha$-process or, equivalently, the
structural relaxation results from a series of MB transitions.
Further, we suggest that the $\beta$-process is related to the
exploration of a MB, i.e., to the E-sequences. In this picture, both
relaxation phenomena are naturally multi-step processes. Further, the
growing timescale separation of the $\alpha$- and the $\beta$-process
upon cooling is a consequence of the increasingly important trapping
of the system in a MB. When decreasing the temperature the system is
forced into deeper and deeper regions of the MB so that smaller and
smaller regions of the MB can be explored due to the $\beta$-process
and, consequently, its relaxation strength decreases. In other words,
the number of IS visited during the $\beta$-process is temperature
dependent. Finally, the relative amplitude of the $\alpha$- and the
$\beta$-process depends on the steepness of the MB and the number of
IS combined in the MB. Thus, the experimental observation that the
relative strength of both relaxations is different for various glass
formers can be met. Of course, our speculations need to be carefully
checked in future investigations. As pointed out by Debenedetti and
Stillinger~\cite{Debenedetti}, suitable data is not available at the
present time.

In conclusion, the following picture appears to emerge for IS and MB
dynamics at sufficiently low temperatures. When the system jumps
between IS organized into a MB, groups of spatially correlated
particles achieve comparatively large displacements by performing
string-like motion. In doing so, the strings arise, for the most
part, not due to a coherent motion of all involved particles during a
single IS transition, but from the interplay of displacements of
various sub-units taking place at different times. During the
exploration of the MB, the system enters energetically less favorable
regions near the ridge of the MB from which it either escapes to a
new MB or returns to the bottom of the old MB. While in the latter
case, the formed strings dissolve due to the backward motions of the
involved particles, they persist in the former. The particle
rearrangements during different MB transitions are basically uncorrelated. In
particular, different groups of particles are mobile and form
strings. Hence, one may speculate that successful string-like motion
results in the escape from a MB. We ascribe the exploration of the MB
to the Johari-Goldstein $\beta$-process and transitions between
distinct MB to elementary steps of the $\alpha$-process. In this
picture, both relaxation phenomena are multi-step processes and their
growing time-scale separation on cooling results from the
increasingly important temporary trapping of the system in the MB.

\begin{acknowledgments}

The authors thank Y.\ Gebremichael for stimulating discussions.
M.\ V.\ gratefully acknowledges funding by the Deutsche
Forschungsgemeinschaft (DFG) through the Emmy-Noether Programm.
\end{acknowledgments}

\end{document}